\documentclass[11pt,a4paper]{article}
\newcommand\jhep[3]{{\it JHEP }{\bf #1} (#2) #3}
\newcommand\npa[3]{{\it Nucl. Phys. }{\bf A#1} (#2) #3}
\newcommand\npb[3]{{\it Nucl. Phys. }{\bf B#1} (#2) #3}
\newcommand\plb[3]{{{\it Phys. Lett. }{\bf B#1} (#2) #3}}

\newcommand\prd[3]{{{\it Phys. Rev. }{\bf D#1} (#2) #3}}
\newcommand\prep[3]{{{\it Phys. Rep. }{\bf #1} (#2) #3}}
\newcommand\ibid[3]{{\it ibid. }{\bf #1} (#2) #3}
\newcommand\epjc[3]{{\it Eur.\ Phys.\ J. }{\bf C#1} (#2) #3}
\newcommand\zpc[3]{{\it Zeit.\ Phys. }{\bf C#1} (#2) #3}

\newcommand\jetp[3]{{\it Sov. Phys. JETP}{\bf #1} (#2) #3}
\newcommand\sjnp[3]{{\it Sov. J. Nucl. Phys.}{\bf #1} (#2) #3}

\begin{document}

\titlepage

\begin{flushright}
IPPP/02/24\\
DCPT/02/48\\
\end{flushright}

\vspace*{2cm}

\begin{center}
{\Large \bf Multiparticle Production in QCD Jets} \\
 \vspace*{1cm}
J.R.~Andersen$^a$, N.H.~Brook$^b$, Yu.L.~Dokshitzer$^c$, V.A.~Khoze$^a$, W.~Kittel$^d$,
W.~Ochs$^e$, W.J.~Stirling$^a$ and G.~Zanderighi$^a$ \\

\vspace*{0.5cm} {\it $^a$ IPPP, Department of Physics, University of Durham, Durham DH1~3LE, UK \\
$^b$ H.H.~Wills Physics Laboratory, University of Bristol, Bristol BS8~1TL, UK  \\
$^c$ LPTHE, Universit\'{e} Pierre et Marie Curie (Paris VI), 75252 Paris, France  \\
$^d$ HEFIN, University of Nijmegen/NIKHEF, 6515 ED Nijmegen, The Netherlands  \\
$^e$ Max Planck Institut f\"ur Physik, D-80805 Munich, Germany  \\
}
\end{center}

\vspace*{1cm}

\begin{abstract}
We briefly summarise the main results presented at the IPPP Workshop on Multiparticle
Production in QCD Jets, held in Durham in December 2001.
\end{abstract}

\newpage
\renewcommand{\thefootnote}{\fnsymbol{footnote}}
\section{Introduction}
\label{sec:intro}

 During the last few years, particle physics experiments have collected
an impressive amount of new information on multiparticle production in
hadronic jets, which allows many important tests of both perturbative and nonperturbative aspects of
QCD to be performed. It is therefore timely to survey the current
status of such studies and to consider future directions for work in this field.

The main aim of the IPPP Workshop on ``Multiparticle Production in QCD Jets"
was to survey recent theoretical results and the experimental data from the LEP era,
HERA and the Tevatron. Particular attention was paid to the prospects for QCD studies at the LHC and
at a future linear $e^+e^-$ collider.


\section{Jet Production and Fragmentation\protect\footnote{Author: N.~Brook}}
\label{sec:jet-prod-fragm}

The D0~\cite{varelas} and CDF~\cite{korytov}
 collaborations presented results on jet fragmentation at the
 Tevatron,  in particular on the momenta spectra of charged particle distributions
and the subjet multiplicities.

The CDF collaboration have investigated~\cite{CDF}
 the momenta spectra of charged particles in
central dijet (or $\gamma-{\rm jet}$) events with a dijet mass in the range
$ 80 < M_{jj} < 630\ {\rm GeV}.$ To limit any biases being introduced,
due to the fact that there is underlying event debris in the hadron collider
environment, the analysis is confined to small opening angles around the
jet axis.
The scaled momentum
distribution of the charged tracks, $\xi=\log(E_{\rm jet}/p),$ associated
with a jet was
measured and compared to the  Modified Leading Logarithmic Approximation (MLLA) 
 parton predictions~\cite{MLLA}. The parton
predictions are related to the hadron distributions, according to the
Local Parton  Hadron Duality (LPHD) picture~\cite{LPHD}, by a simple
normalisation factor, $K.$ In the MLLA theory
the dependence of the $\xi$ distribution
on the jet energy and cone angle, $\theta,$ is given by the
variable $Y=\log(E_{\rm jet}\sin\theta/Q_{\rm eff}),$ where $Q_{\rm eff}$
is a cut-off scale for the gluon emissions. The CDF fits to the data
confirm the scaling of the momenta spectra with $E_{\rm jet}\sin\theta$
with a $Q_{\rm eff} = 240\pm 40 {\rm\ MeV}.$

The CDF collaboration have also studied the evolution of the charged particle
multiplicity with $M_{jj}.$ Using the value of $Q_{\rm eff}$ obtained above
from the momenta spectra and the fraction of gluon jets in the dijet sample
extracted from the HERWIG Monte Carlo generator~\cite{HERWIG}, it is possible
to extract the value of $K$ and the ratio of parton multiplicities in gluon
and quark jets, $r.$ The fit yielded a value of $K=0.57\pm0.06\pm 0.09$ and
$r=1.7\pm0.3.$ The choice of parton density functions (which affects the
fraction of gluon jets) and $Q_{\rm eff}$ have negligible effect on the value
of $r$ but are a large contribution to the systematic errors on $K.$ This
value of $K$ only reflects the ratio of the number of charged particles to
the number of partons and is consistent with the LPHD hypothesis of a
one-to-one correspondence between final partons and all observed hadrons.
The value of $r$ is also consistent with that expected from MLLA.

The D0 collaboration have also studied the internal jet structure through
resolving subjets~\cite{D0}. The motivation was again to study the difference
between quark and gluon jets. To achieve this it was necessary to select
enriched quark and gluon jet samples. This was realised by probing the
internal structure of jets at the same $E_T$ but from $\rm p \bar p$
collisions at
different centre of mass energies ($\sqrt{s} = 630$ and $1800 {\rm\ GeV}$).
For jets with $55 < p_T({\rm jet})< 100 {\rm\ GeV}$ a
gluon enriched sample is found at the higher $\sqrt{s}$ compared to a quark
fraction that is enhanced at the lower $\sqrt{s}.$ The relative fraction
of quark/gluon events is estimated using the HERWIG Monte Carlo
generator.

The $k_T$ jet algorithm~\cite{ktalgo}
 was used in this analysis; in addition to the
$p_T$ criteria listed above, the jets were required to be centrally produced, i.e. 
have a pseudorapidity $|\eta| < 0.5.$  The subjet multiplicity was calculated
by using the ``pre-clusters" associated with a particular jet and
re-applying the jet algorithm and using a jet resolution
parameter equivalent to the minimum subjet
$p_T$ being $\sim 3\%$
of the total jet $p_T.$ The average subjet multiplicities
were measured to be $M_{1800}=2.74\pm0.01$ and $M_{630}=2.54\pm0.03$ at
$\sqrt{s} = 1800 {\rm\ and\ } 630 {\rm\ GeV}$ respectively. Applying
corrections for detector effects and the fraction of gluon events at
each $\sqrt{s}$, the mean subjet multiplicity of a quark jet, $M_q=1.69\pm0.04,$
and a gluon jet, $M_g=2.21\pm0.03,$ can be derived. Furthermore from the
fact that $M-1$ corresponds to the average number of subjet emissions a
value of $r$ can be extracted again. D0 measured
$r=1.84\pm0.15\pm^{0.22}_{0.18}$ with the major systematic error being
the determination of the gluon jet fraction. This complementary
measurement of $r$ is consistent with the value presented by CDF.

The motivation for limiting the cone size in the CDF analysis of the
charged particle distibution was the need to understand the so-called
underlying event. The underlying event is defined as everything not
associated with the hard two-to-two subprocess. Experimentally the
underlying event is measured using minimum bias data -- soft inelastic
collisions in which both the incoming hadrons break up.
It has been shown~\cite{field1}
 that the underlying event  in hard scattering
has considerably more activity than the soft collisions for the same
available energy. The model currently implemented in the HERWIG Monte Carlo
event generator is based on the minimum bias generator of UA5 and fails
 to reproduce this additional activity.

Borozan {\it et al.}~\cite{borozan}
have been developing a model for soft interactions
which is founded  on a firm theoretical basis.
The model is based on an eikonal approach. It takes into account
contributions from hard multiparton interactions and a soft component,
representing
the opaqueness of the hadron as a function of the available energy and
impact parameter. This was shown to successfully describe the CDF data at
$\sqrt{s}=1800 {\rm\ GeV}.$

At Tevatron energies it is known that the fixed order QCD calculations for
heavy quark production underestimate the cross section for $b$-production
when compared with the data. Another approach~\cite{sjostrand}
 to heavy quark production is that of
parton showers (PS). Although only approximate, it has the advantages of
resumming large logarithms, being process generic and easy to match to
hadronisation. The PS approach convolutes the two-to-two hard scatter
with initial state and final state (QCD) radiation to represent the
two-to-many final parton configuration. In the PS approach, 3 main
contributions to heavy quark production can be isolated:
\begin{enumerate}
\item pair creation ($gg\rightarrow Q \bar Q$  and $q\bar q
\rightarrow Q \bar Q$ plus additional showering)
\item flavour excitation based on the $c$ and $b$ content of the parton
density functions ($Qq \rightarrow Qq$ and $ Qg \rightarrow Qg$
\item gluon splitting (e.g. $gg \rightarrow gg$ with $g \rightarrow Q
\bar Q.$)
\end{enumerate}

By including the contributions to $b$-quark production from the PS it is possible
to describe the $b$-quark cross sections in the central rapidity regions as a
function of the minimum transverse momentum of the $b$-quark at
CDF~\cite{field2}.

The hadronisation of heavy quarks should not be studied in isolation
from the production environment. There are ``beam" remnant issues that
affect the $b-{\rm quark}$ distributions due to ``drag" effects. In the
string model there are 3 hadronisation mechanisms:
\begin{enumerate}
\item normal string fragmentation leading to a continuum of phase-space
states,
\item cluster decay, where a low mass string leads to an exclusive
two-body state, and
\item cluster collapse, where a very low mass string becomes a single
hadron.
\end{enumerate}
These production mechanisms can lead to a production asymmetry between
hadrons containing heavy quarks and antiquarks. These production
asymmetries have been observed in fixed-target experiments studying
charm production~\cite{fixed}. The modelling of the low-mass clusters is
particularly sensitive to these asymmetries and the low-mass behaviour
of the models has been refined~\cite{norrbin} in order to reproduce the effect
observed by experiment. The phenomenological parameters introduced into
the framework leave a large leeway in the predictive ability for these
asymmetries for $b$-quarks at higher centre-of-mass collisions.
It is important that the study of heavy quark production is made
 in conjunction with the  development of the multiple interaction models.

By studying high energy cosmic rays it is hoped to gain insight into their
source and acceleration mechanism. Particular features that are under study
are the `knee' in the particle flux spectrum that occurs at $\sim 10^{3}
{\rm\ TeV}$ and the precise form of the spectrum at energies $>10^7
{\rm\ TeV.}$ To study cosmic rays at and above these energies it is
necessary to understand the properties of the particle showers that are
produced as the
cosmic ray enters the atmosphere. The mass and energy of the
primary particles are deduced from the properties of these air showers.
The form of the air showers are dependent on the physics of the hadronic
interactions, electromagnetic interactions, particle production, decays
and transport mechanisms.  The reconstruction of the primary energy and mass is
particularly sensitive to the height of the shower maximum,
the lateral particle distribution and particle
content of the air showers at ground level. The ability to reconstruct
the mass is currently limited by the modelling of hadronic and nuclear
interactions~\cite{knapp}.
The hadronic models being developed
are based on Gribov-Regge theory of multi-Pomeron exchange for the
dominant soft interactions. Particle production is modelled
including string
fragmentation. They have to take into account extrapolation to higher
energies, diffraction, hard processes and nuclear
interactions. The models are beginning to describe the cosmic ray data
rather well. The various models available converge: for example the
differences in the cross section, $\sigma_{\rm p-air},$ are within 8\%
and the shower maximum within $6 {\rm\ g/cm^2}.$ Unfortunately there are
still areas where the uncertainties are greater than 10\% where it is
hoped that future accelerator-based HEP experiments, such as those based at heavy ion facilities,
could help in increasing the understanding of the processes involved.
The Auger experiment, under construction in Argentina, will soon allow
investigation of the hadronic interaction models at energies as high as
$10^8 {\rm\ TeV.}$

\section{Perturbative and Non-perturbative Aspects of Multiparticle
Production\protect\footnote{Author: W.~Ochs}}\label{sec_Ochs}

The production of multiparticle final states in hard collisions
according to the standard description involves three phases: first there is
a hard primary interaction at small distance scales
with few outgoing partons, leptons or gauge bosons, subsequently
the produced partons evolve into parton jets
according to a perturbative description,
and finally, there is a non-perturbative
transition into the observable hadrons at large distances. All three phases
of particle production have been addressed at the workshop.
 The aim is the improvement of the theoretical description,
calculations of new
observables and the experimental tests of predictions.

\subsection{Improved treatment of primary hard processes}
An improved accuracy is required for a better test of the theory
and also for a better understanding of QCD processes in the search
for new physics at future colliders. The parton shower calculations
as applied in event generators are accurate only in the phase space regions
near the soft and collinear singularities but typically fail at large angles
where full matrix element results are needed.

For the case of  $e^+e^-$ annihilation an algorithm is
presented by L\"onn\-blad \cite{Lonnblad:2001iq} in which the
Colour-Dipole Cascade Model, as implemented in the \textsc{Ariadne} program, is
corrected to match the fixed order tree-level matrix elements for
$e^+e^-\rightarrow n$ jets. The result is a full parton-level
generator for $e^+e^-$ annihilation where the generated states are
correct at tree-level to fixed order in $\alpha_S$ \emph{and} to all
orders with MLLA accuracy. In addition,
virtual corrections are taken into account to all orders with MLLA
accuracy.
Results have been presented and discussed, where matrix elements are
used up to second order in $\alpha_S$.
The algorithm is applicable also for higher orders and it should be
possible to modify it to work even for hadronic collisions.

Similar problems are met in the computation of hadronic electroweak gauge
boson pair production ($\gamma\gamma,\;  ZZ,\; WW,\; Z\gamma, \; W\gamma,\;
WZ$) as discussed by
Burby \cite{burby}. These processes are of interest, in particular in the
search for the Higgs boson and physics
beyond the Standard Model.
A complete framework for
analyses is provided by parton shower based event generators.
It is demonstrated how
the description of the
high $p_T$ emissions in gauge boson pair production
is improved by merging the QCD
matrix elements with the parton shower algorithm of
\textsc{Pythia}.
A method of overpopulating the
shower phase space and then rejecting events down to the matrix element rate is
employed by calculation of a correction factor which takes into account the
possible parton shower histories. In addition to
gluons, real photons are permitted to shower and hence must also be
included in the matrix element corrections.

Another problem which occurs in the calculation of radiative QCD corrections to pair
production of $W$ and $Z$ bosons,  as well as to
 top quarks, is the finite width of these particles. This has been discussed by
Chapovsky \cite{Chapovsky}. The structure of higher-order radiative corrections for
processes with unstable particles is analysed. The mass $M$ and width $\Gamma$ of the
unstable particles yield a hierarchy of scales. By subsequently integrating out the
modes induced by these scales, a hierarchy of effective field theories is obtained.  In
the effective field theory framework the separation of physically different effects is
achieved naturally. In particular, a separation of factorisable and non-factorisable
corrections to all orders in perturbation theory is obtained. The non-factorisable
corrections can be classified in two ways. First of all, they can be of the
production-decay and decay-decay types. Second, they can be due to interaction with the
production/decay dipoles and propagation corrections. It is known that one-loop
non-factorisable corrections to invariant mass distributions are suppressed by powers
of $M/E$ at high
energies. The mechanism of this suppression is studied
and estimates of higher-order
non-factorisable corrections at high energy obtained.
These results are applied to the above pair production processes.

\subsection{Semisoft processes: experimental tests of perturbative
predictions}

Various details of the multiparticle final state
concerning particle multiplicities and their inclusive differential
distributions
have been studied in the recent years and analytical results derived.
This involves particles with energies around 1 GeV or smaller.
A simple model relates the partonic distributions obtained after an
evolution of the parton cascade towards small scales ($Q_0\sim \Lambda$)
directly to the corresponding hadronic distributions. Such a correspondence
could hold in a dual sense and was originally proposed for single inclusive
spectra (LPHD \cite{LPHD}). Detailed experimental data
became available from LEP
(for a review see \cite{ob}) and more recently from HERA. The status of the model 
was reviewed by Ochs (see also \cite{ko}).

It has been concluded that the agreement between  the perturbative predictions
and experiment had been  generally successful,  depending on the accuracy of
the calculation. Global quantities like mean multiplicities and
correlation moments can be described at the quantitative level.
The characteristic features of
single inclusive particle distributions (`hump backed plateau') are well
established for all types of particles and the approach to asymptotic
scaling (`$\zeta$-scaling') follows the DLA (Double Logarithmic Approximation) 
and MLLA predictions. However the behaviour in the very soft region
($k_T\sim Q_0$) depends on mass effects and  additional model
assumptions are needed to obtaim a good description.
Nevertheless, the early onset of asymptotic behaviour of this ultrasoft
region is generic and its verification supports the underlying
soft gluon coherence and `angular ordering'. Angular correlations agree
with the various asymptotic (DLA) predictions at least qualitatively
and with numerical results at the almost quantitative level.
Calculations concerning the transition $y_{cut}\to 0$
from jet to hadron and the transition to quasi-exclusive processes involving
large rapidity gaps have been surprisingly successful. A problem has
recently been met with multi-correlations of soft ($k_T\sim Q_0$) particles
(see below).

In most applications of this type the relative momenta can become small
and therefore the coupling $\alpha_s(k_T)$ can become large of order 1. The extension
of the calculations into this kinematic region is not justified a priori,
but the perturbative expansion is converging rapidly and one can consider
the overall successful description at least as an effective
parametrisation of the soft part. Remarkably, the  coupling
$\alpha_s$ obtained at small scales is roughly consistent with the experimental
results for the integral
$1/\mu\int_0^\mu \alpha_s(k)dk $ obtained from fits to shape variables using
the perturbative approach with power corrections added 
(see Section~\ref{sec:dokshitzers-writeup} below).

Global moments of multiplicity distributions of quark and gluon jets, in
particular their mean values $N_{F,G}$, have been computed by solving the
 so-called MLLA evolution equations, as discussed by Dremin
\cite{id} (for a recent review \cite{dg}). At higher energies the solutions can be written
as  $N_{F,G}\propto \exp(\int^y \gamma_{F,G}(y')dy')$ with the anomalous
dimensions $\gamma_{F,G}$   calculated in powers of
$\gamma_0\propto\sqrt{\alpha_S}$, where the LO and NLO terms for $N_F$ and
$N_G$ are the same. The ratio
 $r=N_G/N_F$ has been calculated to order $\gamma_0^3$, which corresponds to
the inclusion of the  4NLO expansion of $\gamma_{G,F}$. Asymptotically $r\to
C_A/C_F=9/4$, and  at LEP energies with increasing accuracy
the calculations approach
the data and the
exact numerical solutions of the evolution equations finally 
give good agreement.
It is advantageous to
study the quantities
$r^{(1)}=N_G'/N_F'$ and $r^{(2)}=N_G''/N_F''$
as they have small higher order corrections. QCD predicts
\begin{equation}
r<r^{(1)}< r^{(2)}<C_A/C_F=2.25.\label{r1r2}
\end{equation}

The experimental determinations of the ratio $C_A/C_F$ from
jet multiplicities and from scaling violations of particle spectra
by DELPHI have been
explained by Siebel \cite{delphi}. The energy dependence of the
jet multiplicities  $N_F$ and  $N_G$
is determined from quark and gluon jets isolated in
$e^+e^-\to$ 3 jets using certain jet algorithms. The above ratio
$r^{(1)}$ has been determined this way and the results are
less sensitive to non-perturbative effects than the absolute
values of the multiplicities. The relation (\ref{r1r2}) has been confirmed by this 
procedure.

In an alternative approach, the total multiplicity of 3-jet events as a
function of jet angles is compared to the theoretical description which does
not assign particles to individual jets but rather takes into account the coherent
emission from the $q\bar q g$ `antenna'. The results from both methods
are consistent with the QCD expectation $C_A/C_F=9/4$.

The `classical' MLLA prediction of the hump-backed plateau, namely the
Gaussian type distribution of particles in the variable $\xi=\log(1/x_p)$
(with $x_p=p/p_{max}$ determined in the current region of the Breit frame)
has been scrutinised by Brook using data from LEP and HERA \cite{brook}.
The moments have been extracted from fits of a distorted Gaussian to data
around the maximum and compared to several theoretical predictions
which differ in their treatment of soft particles (large $\xi$).
The data are bracketed by the MLLA predictions with and without
inclusion of a mass effect (taking $Q_0$ as the particle mass).
A good description can be obtained by
introducing an additional effective mass parameter. At high energies
all descriptions converge and agree with the data. It should be recalled
here, that the MLLA description
is not very good for soft particles because of slow convergence of the
$\sqrt{\alpha_S}$ expansion, and better results can be obtained by using the first two
terms of the perturbative expansion.

The recent results from LEP and HERA on multiparticle angular
correlations have been discussed by Chekanov.
In general, two-particle densities and angular correlations
  in DIS jets agree  with the DLA calculations \cite{zeus1}. 
The normalised factorial moments, measured in
  rings around the jet direction, show discrepancies
  with the perturbative QCD predictions for all three experiments,
  L3 \cite{L3}, DELPHI \cite{delphi2} and ZEUS \cite{zeus2}. From
  parton-level Monte-Carlo studies, it was concluded that these
discrepancies are related to the approximations inherent in the  analytical DLA
results (especially neglect of energy-conservation constraints).
A more serious discrepancy with the theoretical predictions has been found recently for
correlations in   restricted $p_t$ and $p$ regions by ZEUS\cite{zeus2}.
 This is the  first example where perturbative QCD,
  in conjunction with the LPHD hypothesis, fails  on a qualitative level.
 As in case of the single particle spectra, the
problem is with particles  near the non-perturbative boundary $p_t\sim Q_0$
where mass effects become important.
It is suggested elsewhere that the predictions should work for mini-jets with
low virtuality around $Q_c\sim1$~GeV.

\subsection{Non-perturbative models}
At HERA and the TEVATRON sizeable fractions (10\% and 1\% respectively) of events with two
high $E_T$ jets have a large rapidity gap. A possible explanation of such events is the
presence of a hard colour singlet exchange. Metzger \cite{metzger} reported on
an analysis of $e^+e^-\to q\bar q g$ events from hadronic $Z$ decays obtained by the L3
collaboration to search for additional evidence for the presence of this mechanism.

In the JETSET parton shower model of this process,
hadrons are produced from two
strings connecting the gluon with the quark and the antiquark.
Models with colour singlet exchange are constructed  in which a
string is built up between the $q\bar q$ pair, whereas the gluon
is disconnected from the quarks and in one example hadronises like a
boosted $q\bar q$ pair.
Particle flows between the three jets are found to be sensitive to the
presence of the colour singlet  mechanism.
The result of this study is that the
data can accommodate no more than about 8\% colour-singlet exchange (at 95\%
CL). Furthermore, the data are found to be incompatible with the recently
proposed model by Rathsman assuming a colour connection
based upon a generalised Lund area scheme. 
Similar methods are applied to four-jet events in $e^+e^-\to W^+W^-$ to place
limits on various colour reconnection models.

The possibility of another non-perturbative phenomenon in jet evolution,
squeezed states, was
considered by Kuvshinov \cite{kuvshinov}. This phenomenon
is known in quantum optics and can
be obtained from coherent states. The time evolution of a gluon coherent
state is studied using a QCD Hamiltonian approach.
It is shown that after a small period of time, squeezed states appear due to
the self-interaction of the gluon quantum.
The gluon correlation function has singularities at very small angles.
Furthermore, there are connections between squeezing and a chaotic behaviour
of gluons in the jet.


\section{Correlations and Fluctuations\protect\footnote{Author: W.~Kittel}}
\label{sec:corr-and-fluct}

Analytical QCD calculations at the DLA and/or MLLA level, combined with the LPHD
hypothesis \cite{LPHD}, have proven particularly successful in explaining
single-particle distributions and the energy evolution of their integrals, i.e. the
average multiplicity (see Section~\ref{sec_Ochs}). However, it is in the {\it correlations}
between particles that the detailed information on the dynamics underlying their production
may be expected to manifest itself.

High-order correlations have indeed been observed in all types of
particle collisions at high energies in the form of super-Poissonian
multiplicity fluctuations in small regions of phase space \cite{1}.
With improving resolution, these fluctuations exhibit an approximate power-law
scaling analogous to that observed for fractal objects, where the power is
directly related to the anomalous dimension. The order dependence of this
anomalous dimensions leads to the conclusion that particle production
in a QCD jet follows a self-similar multifractal pattern, as in fact expected
for a multiplicative branching process like QCD showering \cite{2}.
Deviations from exact scaling exist, however, in the form  of  saturation
at small virtuality, due to the running of the coupling constant.

Detailed QCD predictions exist on the parton level for the so-called
$H_q$ moments \cite{3} measuring the relative amount of genuine $q$-particle
correlation, on the one hand, and for the scaling behavior in angular \cite{4}
and (transverse) momentum variables \cite{5}, on the other.
Assuming LPHD to also hold for these correlations, the calculations have been
tested on the particle level in the L3 \cite{L3,6}, DELPHI \cite{delphi2} and
ZEUS \cite{zeus2} experiments. While the JETSET Monte Carlo model perfectly
reproduces the features of the data, the analytical calculations have to go
as far as NNLA to qualitatively obtain the oscillation pattern of the $H_q$
moments observed for increasing rank $q$. This and the failure to reproduce
the scaling behavior in angular variables may be explained by the various
approximations required in the analytical calculations, in particular the
lack of energy-momentum conservation at each branching.  However, the limit
of validity of LPHD is reached when trying to reproduce the fluctuation
pattern as a function of the limit on the (transverse) momentum.  Since the
analytical calculations are for partons at asymptotic energies, the
predictions will be more readily tested at LHC energies.


\section{Jet Shape Observables\protect\footnote{Authors: Yu.L.~Dokshitzer and
G.~Zanderighi}} \label{sec:dokshitzers-writeup}

Over the last few years the theory and phenomenology of jet shape observables
has attracted a lot of attention.
Nowadays, when  jet shapes in $e^+e^-$ annihilation have been analysed
theoretically, both perturbatively~\cite{PTstandard} and at the level
of the leading non-perturbative $1/Q$ power
corrections~\cite{NPstandard}, the focus has shifted to the more complicated
environment of DIS
~\cite{TDIS,BDIS,kout-dis} and hadron-hadron
collisions~\cite{kout-hh}.

In particular, new results have been obtained by the Milan group  for the out-of-plane
QCD radiation in  three-jet events in DIS~\cite{kout-dis} and in hadronic
collisions~\cite{kout-hh}, and were presented at the Workshop by Banfi and Smye respectively. The
perturbative analyses have been carried out for the first time in hadronic collisions
involving more than two emitting jets, at the next-to-leading (NLL; single-logarithmic,
SL) accuracy. The novel feature of the power-suppressed non-perturbative corrections to
observables, involving hadrons in the initial state, is their dependence on the initial
parton distributions, which enters at the SL level.  As discussed by Dasgupta, this
same feature has been found for the first time in the study of jet broadening in
DIS~\cite{BDIS}, whose properties were already known to be subtle in the simpler $e^+e^-$
annihilation case~\cite{broad-np}.

Recently the question about the actual need for non-perturbative $1/Q$
power corrections has been raised by Hamacher.  He has
demonstrated that, at least according to the DELPHI analysis~\cite{hamacher}, a
broad set of mean values of jet characteristics can be consistently
described within the so-called renormalisation-group improved (RGI) scheme for
dealing with the perturbative series, without invoking any explicit $1/Q$
terms.
However the RGI-perturbative approach itself has  limited power, as it applies only to
the {\em mean values}\/ and only to $e^+e^-$ annihilation, i.e. to
 one-parameter problems.
In particular, it does not apply either to event shape distributions or the mean values in
DIS and/or hadron-hadron collisions. Nevertheless, the amazing accuracy
of the RGI-perturbative treatment of the $e^+e^-$ mean shape observables poses a
challenge for theorists: why does this specific prescription of
guessing the higher-order perturbative terms seem to be so much more
reliable than the other schemes.

The determination of resummed distributions of event shape variables
and jet-rates has up to now proved very labour-intensive. This is
despite the fact that many aspects of the calculations are actually
fairly similar from one observable to another -- for example one can
usually make the same approximation concerning the multi-particle matrix
elements and also exploit the same factorisation techniques.
Zanderighi described an attempt at
a numerical, completely general
approach~\cite{numsum}.
The main idea of this approach is to relegate a hard but routine theoretical analysis
(all order resummation of double and single logarithms, matching with the exact matrix
element calculation, etc.) of, virtually, all arbitrary collinear- and infrared-safe
jet observables, to numerical calculation on a computer.

\section{BFKL Physics\protect\footnote{Authors: J.R.~Andersen and
W.J.~Stirling}}\label{sec:bfkl}

One of the most interesting and challenging frontiers of perturbative QCD is
provided by hadronic scattering processes in the so-called high-energy limit,
i.e. asymptotically large centre-of-mass scattering energy but fixed and finite
momentum transfer. In QCD, the description of this regime is provided by the
BFKL approach \cite{bfkl}. Although a relatively new field, there has been
remarkable progress in both theoretical understanding and calculations
and in experimental tests.

\subsection{BFKL at NLO: status and outlook} \label{sec:fadins-talk}
Fadin \cite{fadin} gave an overview of the BFKL approach to QCD processes in the high-energy limit,
highlighting the achievements to date and some outstanding issues.
The approach is based on one of the
remarkable properties of QCD -- gluon Reggeisation --  which is extremely important
for high energy QCD. The Pomeron, which determines the high energy behaviour of the
cross sections, and the Odderon, responsible for the difference of particle and
antiparticle cross sections, appear in this approach as compound states of two and
three Reggeised gluons respectively. It is valid not only in the leading logarithmic
approximation (LLA), where the BFKL approach was first developed, but also in the
next-to-leading approximation (NLA). The scattering amplitude for the high energy
process $A+B\rightarrow A^{\prime }+B^{\prime }$ at fixed momentum transfer $\sqrt{-t}$
in the NLA, as well as in LLA, is given by the convolution of impact factors $\Phi
_{A^{\prime}A}$ and $\Phi _{B^{\prime }B}$, that describe the transitions $A\rightarrow
A^{\prime }$ and $B\rightarrow B^{\prime }$ and the Green's function for the two
interacting Reggeised gluons. Both the kernel of the integral equation satisfied by the
Green's function and the impact factors are unambiguously defined in terms of the gluon
Regge trajectory and the effective vertices for the Reggeon-particle interactions,
which have been calculated at next-to-leading order (NLO) in a series of papers. 
The explicit form of the NLO kernel for forward scattering (with
colourless exchange in the $t$ channel) was found.  The cancellation of the infrared
singularities in the kernel and in the impact factors of physical (colourless)
particles was demonstrated. At the parton level (for quarks and gluons) impact factors
were calculated in the NLO approximation for arbitrary momentum transfer $t$ and any
colour state in the $t$-channel. For the non-forward kernel the only piece which is not
yet calculated is the contribution of the two-gluon production for the colour singlet
in the $t$-channel. The colour octet kernel is found for arbitrary $t$. The
compatibility of the gluon Reggeisation with the $s$-channel unitarity (``bootstrap''
of the gluon Reggeisation) was proved.

Two important problems remain unsolved. One is the calculation of the only missing
piece in the non-forward kernel -- the contribution of two-gluon production; the
second is  the calculation of the impact factors for highly virtual photons. The
knowledge of the non-forward kernel means a significant extension of the region of
possible phenomenological applications. The virtual photon impact factors play a
specific role since they can be calculated ``from first principles'' and since there
are experimental data for the high energy $\gamma^*\gamma^*\to$hadrons  cross section.

The calculation of the NLO BFKL kernel revealed a new problem -- the large size of the
scale-invariant correction to the eigenvalue function of the kernel, and also initiated
a new wave of papers devoted to an old one -- the running of the coupling. Several ways
of solving the first problem are suggested. Possible ways of including
the running coupling  and its implications have also been  widely discussed.

\subsection{Monte Carlo approach to studying the BFKL chain}
\label{sec:monte-carlo-approach}

Andersen presented results on forward jet and $W$ production at hadron colliders
obtained using a BFKL Monte Carlo model first reported in Ref.~\cite{Orr:1997im}. The
basic idea of the Monte Carlo BFKL model is to solve the BFKL equation while
maintaining kinematic information on each radiated gluon. This is done by unfolding the
integration over the rapidity-ordered BFKL gluon phase space and introducing a
resolution scale $\mu$ to discriminate between resolved and unresolved radiation. The
latter combines with virtual corrections to form an IR safe integral. Thus the solution
to the BFKL equation is recast in terms of phase space integrals for resolved gluon
emissions, with form factors representing the net effect of unresolved and virtual
emissions.

The benefit of using the BFKL MC is that it takes account of both energy and momentum
conversation in hadronic \lq BFKL' processes and also the inclusion of effects from the
running of the coupling. In contrast, the analytic solution of the BFKL equation
requires the integration over the full rapidity ordered phase space to obtain the
parton level cross section. When it comes to including the parton fluxes from protons
through the parton distribution functions, 
the contribution from the BFKL gluons to the Bjorken $x$'s is
therefore inaccessible in the standard approach. This means that in practice only the
contribution to the subprocess centre-of-mass energy from the leading jets in
(Mueller-Navelet) dijet production with a BFKL gluon ladder exchange can be taken into
account in the standard BFKL approach. This  inevitably leads to an overestimate of the
parton flux. The BFKL MC approach also enables the details of the final state to be
investigated.

The main result of the study \cite{Andersen:2001kt} is that the contribution of the
BFKL gluon radiation to the parton momentum fractions (at LHC energies) lowers the
parton flux in such a way as to approximately cancel the rise in the subprocess cross
section with increasing dijet rapidity separation ($\hat\sigma_{jj} \sim
\exp(\lambda\Delta y$)) predicted from the standard BFKL approach. The leading-order QCD
prediction for the dijet cross section is therefore only slightly modified to an almost
no-change situation compared to leading order QCD.  However, other BFKL signatures such
as the dijet azimuthal angle decorrelation do still survive.

Although at hadron colliders the simplest process exhibiting BFKL behaviour is the
production of dijets with large rapidity separation, the formalism also applies to the
production of more complicated forward final states. In particular, Andersen discussed
the situation where one of the forward Mueller-Navelet jets is replaced by a $W$--jet
pair, which also  provides a testing ground for BFKL signatures~\cite{Andersen:2001ja}.
In fact, the suppressing effect of the BFKL gluon radiation on the pdfs is less
pronounced in this case, since requiring a $W$ in the final state means that at least
one of the initial state partons must be a {\it quark}, with a less steeply falling
pdf. This means that the BFKL rise in the partonic cross section is not compensated to
the same extent as in the dijet case when including the pdfs to calculate the hadronic
cross section.

\subsection{The Linked Dipole Chain model}

Gustafson reported on work with Miu \cite{Gustafson1} on the approach to BFKL
asympototic behaviour in the framework of the
Linked Dipole Chain Model (LDCM) \cite{Gustafson2}. The starting point is the
observation that the small-$x$ structure function $F_2$ measured at
HERA is well parameterised by the form $F_2 \sim c(Q^2) x^{-\lambda(Q^2)}$, which
resembles the asymptotic BFKL form apart from the fact that the effective exponent is
$Q^2$ dependent: $\lambda(Q^2) \propto \ln(Q^2/\Lambda^2)$. The region of validity of
this simple parameterisation is $10^{-4} < x < 10^{-2}$. In the $x\to 0$ limit, the
LDCM exhibits the standard factorising BFKL behaviour, $F_2 \sim x^{-\lambda_{\rm
BFKL}}$, and interpolates smoothly into the standard DGLAP behaviour at higher $x$. In
particular, the model agrees well with the $F_2$ data. The question then is what
features of the LDCM are essential for the observed simple behaviour of the data.

The behaviour is understood in terms of a simple interpolating model for small $x$ based on
the LDCM that has a smooth transition between large $k_T$, where ordered chains dominate, and
small $k_T$ where non-ordered chains are most important. In the LDCM, the possibility to
``go down" in transverse momentum along the chain, from $\kappa = \ln(k_T^2)$ to a smaller value
$\kappa' = \ln({k'_T}^2)$ is suppressed by a factor $\exp{(\kappa - \kappa')}$. The effective allowed distance
$\delta$ for downward $\ln(k_T)$ steps is therefore at most about one unit, and the phase space
factor $\kappa^N/N\!$ is replaced by $(\kappa + N \delta)^N/N\!$. The expression for the structure
function ${\cal F}$ is obtained (for fixed coupling) by combining this with additional
factors of $[\bar\alpha\ln(1/x)]^N/N\!$, and summing over $N$.
  Simple algebra then shows that the DGLAP and BFKL behaviours
are produced  in the large and small $\kappa$ regions respectively. Further details and numerical
studies can be found in Ref.~\cite{Gustafson1}.

\subsection{QCD and LHC: the experimental viewpoint}

Although the primary physics goal of the LHC is to search for the Higgs boson and signs
of physics beyond the Standard Model, there will also be the opportunity to perform
many high-precision, high-energy tests of QCD. A comprehensive overview of the machine, the detectors
and the physics was given by Tapprogge.

The ATLAS and CMS general purpose detectors (GPDs) have been optimised for
high $p_T$ signatures associated with discovery physics. However, they will at the same time, and
without additional resources, allow precise QCD-related measurements. Indeed, a detailed
understanding of QCD will be of the utmost importance for  the discovery and
measurements of any new physics. With an overall proton-proton collision energy of
14~TeV, the LHC will study the strong interaction in an as yet unexplored kinematic
regime. Through processes involving the production of jets, prompt photons, electroweak
gauge bosons and heavy quark flavours, the strong coupling will be measured up to TeV
scales, partonic structure will be probed at very large $Q^2$ and small $x$, and
all-orders resummation of large logarithms in  `two-scale' kinematic regimes will be
tested.

However in order to exploit the new energy frontier to the full, some extensions to the
detectors will be necessary. In particular, the GPDs have been designed and optimised
for high--$p_T$ signatures in the central region ($\vert \eta\vert < 5$). There is
considerable theoretical interest in ``forward physics", much of which goes beyond the
current reach of the GPDs in that it requires the detection of leading protons,
particles and jets beyond $\vert \eta\vert = 5$, and extended coverage of rapidity
gaps.  With such detector capability, one could measure
\begin{enumerate}
\item the total cross section
\item elastic scattering ($d\sigma/dt, \ \sigma_{el}/\sigma_{tot}, \
\rho$~parameter)
\item diffractive scattering (single, double), e.g. $d^2 \sigma /dt dM_X^2$
\item minimum bias event structure
\item properties of rapidity gaps (survival probability, as tools for new physics
searches)
\item hard diffractive scattering (e.g. leading proton with high $p_T$ jets, $W$ bosons etc.)
\item exclusive production of new heavy states, $p +p\to p+ X + p$, where $M_X$
is measured precisely using the leading proton four momenta
\item two photon physics $p + p \to p + p + (\gamma \gamma \to X)$
\end{enumerate}
The proposed TOTEM \cite{TOTEM} experiment would provide appropriate forward physics
capability, with roman pots for leading proton detection and forward telescopes
covering $3 < \vert \eta \vert < 7$ for inelastic event measurements.

Another very interesting possibility is to identify and measure forward {\sl muons}
with, say,  $\vert \eta \vert > 5$. This would allow the  Drell-Yan cross
section $d^2 \sigma^{\mu\mu} / d M^2 d y$ to be measured at low dimuon mass $M$ and high
rapidity $y$ \cite{deroeck}. Recalling that the parton momentum fractions are sampled (at
leading order) at about  $x_{1,2} = M/\sqrt{s}\; \exp(\pm y)$, this would enable very small
$x$ parton structure to be probed. Designs for a forward ``$\mu$station" -- a silicon-based
detector inside the beam pipe -- have been produced \cite{mustation}.


\begin{thebibliography}{99}



\bibitem{varelas} N. Varelas, ``QCD Results from D0", talk given at this
workshop.
\bibitem{korytov} A. Korytov, ``Jet Fragmentation at CDF", talk given at this
workshop.
\bibitem{CDF} CDF Collab., T.~Affolder {\it et al.}, Phys. Rev. Lett. 87
(2001) 211804.
\bibitem{MLLA} Yu.~Dokshitzer {\it et al.}, Int. J. Mod. Phys. A7 (1992)
1875, Yu.~Dokshitzer {\it et al.}, Z. Phys. C55 (1992) 107.
\bibitem{LPHD} Ya.~Azimov {\it et al.}, Z. Phys. C27 (1985) 65, Ya.~Azimov
{\sl et al.}, Z. Phys. C31 (1986) 213.
\bibitem{HERWIG} G.~Corcella {\it et al.}, JHEP 01 (2001) 010.
\bibitem{D0} D0 Collab., V.~M.~Abazov {\it et al.}, Phys. Rev. D65
(2002) 052008.
\bibitem{ktalgo} S.~Catani {\it et al.}, Nucl. Phys. B406 (1993)
187.
\bibitem{field1} R.~Field, ``The Underlying Event in Hard Scattering
Processes'', CDF/MIN-BIAS/PUBLIC/5746.
\bibitem{borozan} I.~Borozan, ``An Eikonal Model for the Underlying
Event in Hadronic Collisions'', talk given at this workshop.
\bibitem{sjostrand} T.~Sj\"ostrand, ``Production and Hadronization of
Heavy Quarks", talk given at this workshop, E.~Norrbin and
T.~Sj\"ostrand, Eur. Phys. J. C17 (2000) 137.
\bibitem{field2} R.~Field, ``The Sources of $b$-quarks at the Tevatron and
their Correlations", UFIFT-HEP-01-25 [hep-ph/0201111].
\bibitem{fixed} WA82 Collab., M.~Adamovich {\it et al.}, Phys. Lett.
B305 (1993) 402, E769 Collab., G.~A.~Alves {\it et al.}, Phys. Rev.
Lett. 72 (1994) 812, E791 Collab., E.~M.~Aitala {\it et al.}, Phys.
Lett. B371 (1996) 157.
\bibitem{norrbin} E.~Norrbin and T.~Sj\"ostrand, Phys. Lett. B442 (1998)
407.
\bibitem{knapp} J.~Knapp, ``Hadronic Particle Production in $10^8$ TeV
Cosmic Air Showers", talk given at this workshop.

\bibitem{Lonnblad:2001iq} L.~L\"onnblad, ``Correcting the Colour-Dipole
  Cascade Model with Fixed Order Matrix Elements,'' hep-ph/0112284.

\bibitem{burby}
S. Burby, this workshop.

\bibitem{Chapovsky}
A.P.~Chapovsky, V.A.~Khoze, A.~Signer and W.J.~Stirling, hep-ph/0108190.

\bibitem{ob}
O. Biebel, \prep{340}{2001}{165}.

\bibitem{ko}
V.A. Khoze, W. Ochs, J. Wosiek, in ``At the frontier of particle physics --
Handbook of QCD'', ed. M. Shifman (World Scientific, Singapore 2001) p.1101.

\bibitem{id}
I.M. Dremin, ``Some problems of the pQCD jet calculus'', hep-ph/0201187.

\bibitem{dg}
 I.M.  Dremin and  J.W. Gary,
\prep{349}{2001}{301}.

\bibitem{delphi}
M. Siebel, this workshop.

\bibitem{brook}
N.H. Brook and I.O. Skillicorn,
\plb{479}{2000}{173}, {\bf B497} 55 (2001).


\bibitem{zeus1}
ZEUS Collaboration, J.~Breitweg {\it et al.},
\epjc{12}{2000}{53}.

\bibitem{L3}
L3 Collaboration, M.~Acciarri {\it et al.},
\plb{428}{1998}{186}.

\bibitem{delphi2}
DELPHI Collaboration, P.~Abreu {\it et al.},
\plb{457}{1999}{368}.

\bibitem{zeus2}
ZEUS Collaboration, S.~Chekanov {\it et al.}, \plb{510}{2001}{36}.


\bibitem{metzger}
W.J. Metzger, this workshop.

\bibitem{kuvshinov}
V. Kuvshinov, this workshop.
\bibitem{1}
A. Bia\l as and  R. Peschanski, \npb{273}{1986}{703}; \ibid{B308}{1988}{857}; \\
 E.A.De Wolf, I.M. Dremin and W. Kittel, \prep{270}{1996}{1}.

\bibitem{2}
G. Veneziano, Proc. 3$^{rd}$ Workshop on Current Problems in High
Energy Particle Theory, Florence 1979, eds. R. Casalbuoni et al.
(Johns Hopkins  University Press, Baltimore, 1979) p.45.

\bibitem{3}
I.M. Dremin, {\it Physics-Uspekhi} {\bf 37} (1994) 715.

\bibitem{4}
W. Ochs and J. Wosiek, \plb{289}{1992}{159}; \ibid{B304}{1993}{144};\\
 Yu.L. Dokshitzer and I.M. Dremin, \npb{402}{1993}{139}; \\
  Ph. Brax, J.-L. Meunier and  R. Peschanksi, \zpc{62}{1994}{649}.

\bibitem{5}
S. Lupia, W. Ochs and  J. Wosiek, \npb{540}{1999}{405}.

\bibitem{6}
 P. Achard {\it et al.} (L3 Collaboration ), ``Measurement of the charged-particle multiplicity
and inclusive momentum distributions in Z decays at LEP", submitted to {\it Phys. Lett.
B}.



\bibitem{PTstandard}
S.~Catani, L.~Trentadue, G.~Turnock and B.R.~Webber,
  \npb{407}{1993}{3}; \\
S.~Catani, G.~Turnock and B.R.~Webber, \plb{295}{1992}{269};\\
S. Catani and B.R. Webber, \plb{427}{1998}{377}.
Yu.L. Dokshitzer, A. Lucenti, G. Marchesini and G.P. Salam, \jhep{01}{1998}{011}.

\bibitem{NPstandard}
M.~Beneke and V.M.~Braun, \npb {454}{1995}{253};\\
R.~Akhoury and V.I.~Zakharov, \plb{357}{1995}{646};
\npb {465}{1996}{295};\\
G.P. Korchemsky and G. Sterman,
\npb {437}{1995}{415};\\
P.~Nason and M.H.~Seymour, \npb {454}{1995}{291};\\
Yu.L. Dokshitzer, G. Marchesini and  B.R. Webber,
\npb{469}{1996}{93};\\
Yu.L.~Dokshitzer, A.~Lucenti, G.~Marchesini and  G.P.~Salam,
\npb{511}{1998}{396},
erratum
  \ibid{B593}{2001}{729}; \jhep{05}{1998}{003}.

\bibitem{TDIS}
V.~Antonelli, M.~Dasgupta and G.P.~Salam, \jhep{02}{2000}{001}.
\bibitem{BDIS}
M.~Dasgupta and G.P.~Salam, hep-ph/0110213.

\bibitem{kout-dis}
A.~Banfi, G.~Marchesini, G.~Smye and G.~Zanderighi,
\jhep{11}{2001}{066};
A.~Banfi, G.~Marchesini and G.~Smye, in preparation.

\bibitem{kout-hh}
A.~Banfi, G.~Marchesini, G.~Smye and G.~Zanderighi,
\jhep{08}{2001}{047}.



\bibitem{broad-np}
Yu.L.~Dokshitzer, G.~Marchesini and G.P.~Salam, \epjc{3}{1999}{1}.

\bibitem{hamacher}
DELPHI Collaboration, Ralf Reinhardt {\it et al.}, `` A study of the energy evolution
of event shape distributions and their means with
 the DELPHI detector at LEP'', to appear in Eur. Phys. J. C; \\
P.~Abreu, {\it et al.}  (DELPHI Collaboration)
\epjc{14}{2000}{557};\\
P.~Abreu {\it et al.}  (DELPHI Collaboration),
\plb{456}{1999}{322};\\
%
DELPHI Collaboration,
DELPHI 2000-116 CONF 415, July 2000.

\bibitem{numsum}
A. Banfi, G.P. Salam and G. Zanderighi, \jhep{01}{2001}{018}.

\bibitem{bfkl} 
V.S.~Fadin, E.A.~Kuraev and L.N.~Lipatov, \plb{60}{1975}{50}; \\
E.A.~Kuraev, L.N.~Lipatov and V.S.~Fadin,  \jetp{44}{1976}{443}; \\
Ya.Ya.~Balitskii and L.N.~Lipatov, \sjnp{28}{1978}{822}.

\bibitem{fadin} A more complete discussion and a full set of references can be found in: V.~S.~Fadin,
\npa{666}{2000}{155}.

\bibitem{Orr:1997im} L.H.~Orr and W.J.~Stirling,
\prd{56}{1997}{5875}.

\bibitem{Andersen:2001kt}
J.R.~Andersen, V.~Del Duca, S.~Frixione, C.R.~Schmidt and W.J.~Stirling,
\jhep{0102}{2001}{007}.

\bibitem{Andersen:2001ja}
J.R.~Andersen, V.~Del Duca, F.~Maltoni and W.J.~Stirling,
\jhep{0105}{2001}{048}.


\bibitem{Gustafson1}
G.~Gustafson and G.~Miu,
\epjc{23}{2002}{267}.

\bibitem{Gustafson2}
B.~Andersson, G.~Gustafson and J.~Samuelsson,
\npb{467}{1996}{443}; \\
%
H.~Kharraziha and L.~Lonnblad,
\jhep{9803}{1998}{006}; \\
%
B.~Andersson, G.~Gustafson, H.~Kharraziha and J.~Samuelsson,
\zpc{71}{1996}{613}; \\
%
B.~Andersson, G.~Gustafson and H.~Kharraziha,
\prd{57}{1998}{5543}.
%

\bibitem{TOTEM} For more information see the TOTEM web site at
http://totem.web.cern.ch/Totem/

\bibitem{deroeck} See for example A.~De~Roeck, ``Parton Density Measurements at Low $x$
at the LHC", Proc. Workshop on Forward Physics and Luminosity Determination at LHC,
Helsinki, Finland,  eds. K.~Huitu {\it et al.}, World Scientific (2001), p. 86.

\bibitem{mustation} See for example V.P.~Nomokonov, ``The Microstation Concept for
Forward Physics", Proc. Workshop on Forward Physics and Luminosity Determination at
LHC, Helsinki, Finland,  eds. K.~Huitu {\it et al.}, World Scientific (2001), p. 164.

\end{thebibliography}
\end{document}